%% file: WarmGasHardSphere_arxiv.tex
\documentclass[twocolumn, nofootinbib, superscriptaddress]{revtex4-1}  

\usepackage{graphicx}				% Use pdf, png, jpg, or eps§ with pdflatex; use eps in DVI mode
\usepackage{amssymb}
\usepackage{dcolumn}
\usepackage{tabularx}
\usepackage{bm}
\usepackage{amsmath}

\newcommand{\etal}{{\em et~al.\/}}
\newcommand{\beq}{\begin{equation}}
\newcommand{\eeq}{\end{equation}}
\newcommand{\beqnl}{\begin{equation*}}
\newcommand{\eeqnl}{\end{equation*}}
\newcommand{\barr}{\begin{array}}
\newcommand{\earr}{\end{array}}
\newcommand{\vecr}{\mathbf{r}}

\newcommand{\vecP}{\mathbf{P}}

\newcommand{\vecE}{\mathbf{E}}

\newcommand{\vecu}{\mathbf{u}}
\newcommand{\vechu}{\hat{\mathbf{u}}}
\newcommand{\vechz}{\hat{\mathbf{z}}}

\newcommand{\vecv}{\mathbf{v}}

\newcommand{\vecG}{\mathbf{V}}
\newcommand{\vecg}{\mathbf{g}}
\newcommand{\vechg}{\hat{\mathbf{g}}}

\usepackage{color}
\newcommand{\todo}[1]{}

\usepackage{dcolumn}
\newcolumntype{.}{D{.}{.}{-1}}
\newcommand{\mc}[1]{\multicolumn{1}{c}{#1}}
\newcommand{\emptycell}{\multicolumn{1}{c}{}}

\begin{document}

\title{Framework for solutions of the Boltzmann equation for ions of arbitrary mass}

\author{D. A. Konovalov}
\affiliation{Information Technology Academy, James Cook University, Townsville QLD 4811, Australia}

\author{D. G. Cocks}
\affiliation{College of Science and Engineering, James Cook University, Townsville QLD 4811, Australia}
\affiliation{College of Chemical and Physical Sciences, Flinders University, Adelaide SA 5042, Australia}

\author{R. D. White}
\affiliation{College of Science and Engineering, James Cook University, Townsville QLD 4811, Australia}

\date{\today}

\begin{abstract}
  We present a framework for the solution of Boltzmann's
  equation in the swarm limit for arbitrary mass ratio, allowing for solutions
  of electron or ion transport. An arbitrary basis set can be used in the
  framework, which is achieved by using appropriate quadratures to obtain the
  required matrix elements. We demonstrate an implementation using Burnett
  functions and benchmark the calculations using Monte-Carlo simulations. Even
  though the convergence in transport quantities is always good, the particle
  distributions did not always converge, highlighting that simple benchmarks can
  give misleading confidence in a choice of basis. We postulate a different
  basis, which avoids a spherical harmonic expansion, which is better suited to
  strong electric fields or sharp features such as low-energy attachment
  processes.
\end{abstract}

\maketitle

\section{Introduction}

The study of non-equilibrium charged particle transport in gases under the
action of applied electric and magnetic fields finds application in a wide
variety of scientific and technological fields ranging from interaction
cross-section and potentials determination, ion mass spectrometry
\cite{White2001d} atmospheric science \cite{Thorn2009a} and plasma discharges
\cite{Bruggeman2016, Boeuf2013}.
Theoretically, such transport is modelled by kinetic methods involving the
solution of Boltzmann's equation \cite{Mason1988} for the phase-space distribution
function, from which the macroscopic properties can be calculated and compared
to experiment.  In recent times, other techniques such as Monte Carlo \cite{Garcia2011, Salvat1999, Semenenko2003},
particle in cell (PIC) \cite{Markosyan2015a} and hybrid/fluid models have been
favoured.  Important however is the need to benchmark these techniques against
other established techniques.

Despite the commonality in the governing equation describing electron and ion
transport in gases -- the Boltzmann equation -- there has been a bifurcation in
the theoretical and computational treatment of electrons and ions.  For
electrons, the smallness in the ratio of the mass of the electron to the neutral
($m/m_0$), has facilitated approximations to the Boltzmann equation. This is
because the velocity distribution function $f(\vecv)$ of electrons undergoing
mostly elastic collisions, is often spherically symmetric and can be represented
well by the first two terms in a spherical harmonic expansion -- the so called
`two-term' approximation\footnote{The two-term approximation for electrons in
  gases can fail when there are inelastic processes.}. The
integro-partial differential nature of the Boltzmann equation can then be
transformed into a second order differential equation facilitating analytic
solution.  For ion swarms however, in general no such approximation can be made,
even for elastic collisions, and the full Boltzmann equation must be solved\footnote{We note that there are theories in the opposing limit $m/m_0\gg 1$ and
  simplified collision operators such as the BGK model.}.  The bifurcation in
the techniques for electron and ion transport then followed, with each
specialising in different mass ratio regime.

This article aims to connect these different mass ratio regimes by presenting a
framework for the solution of Boltzmann's equation in the swarm limit with
arbitrary mass ratios and a flexible choice of basis functions. This is made
possible by directly applying suitable quadratures for the integrals over the
basis functions, which allows one to avoid a spherical harmonic
expansion. In this article, we demonstrate an implementation of the framework
for the Burnett function expansion in spherical harmonics, traditionally used in
electron transport, and apply it to a benchmark hardsphere model. Although this
expansion is well-suited to obtain converged transport quantities we show that
convergence of the distribution functions for high fields and/or high
mass-ratios can require a prohibitively large size of basis. This suggests that
although a traditional benchmark test of the Burnett function expansion can
successful match benchmark values, it will likely fail when ``sharp'' features
such as low-energy attachment processes are present. Furthermore, the asymptotic
behaviour of distributions for ``soft'' potentials are to be
poorly represented by Burnett functions.

We begin with a discussion of common approaches to find solutions to the
Boltzmann equation in section~\ref{sec:background}. The general theory,
including the unified expansion and our demonstration of a basis choice of
Burnett functions, is described in section~\ref{sec:theory}. We then present
benchmark results for several different mass ratios and field strengths in
section~\ref{sec:results} to show the rapid convergence of transport quantities
and draw attention in particular to the convergence behaviour of the
distribution functions. We conclude with some remarks about using different
sets of basis functions in the framework as a means to obtain quicker convergence
and better accuracy in different physical situations.

\section{Background}
\label{sec:background}

The modern era of unified transport theories for electrons and ions started with
firstly the works of Kumar \cite{Kumar1966a, Kumar1967} who reformulated the (near
equilibrium) Chapman-Enskog solution using irreducible tensors and methods
traditionally associated with quantum mechanics and nuclear physics.  These
methods were later adapted to larger field strength non-equilibrium conditions
.  Concurrently,
Viehland, Mason and collaborators formulated the first strong field solution of
Boltzmann's equation for ion transport. which was later applied to
solutions for electron transport \cite{Viehland1975}.
A detailed history is available in several recent reviews on these developments
\cite{Mason1988}.

The general prescription for unified treatments of ions and electrons is the use
of a complete set of Burnett functions to expand the velocity dependence of the
phase-space distribution function, which dates back to the works of Chapman and
Cowling. The Burnett function basis includes a spherical harmonic expansion of
the angular part of the velocity, upon which the associated Lageurre polynomials
with a Maxwellian weighting function are used to expand in the magntiude of the
velocity.  Their utility lies in that these orthogonal functions are themselves
eigenfunctions of the Boltzmann collision operator for a point-charged-induced
dipole interaction.  The computational efficiency that can be achieved with such
a basis set lies in the choice of the Maxwellian weighting function, or
equivalently the zeroth order approximation to the velocity distribution
function.  We now give a brief history of the various `theories' which have been
formulated through intuitive modifications to the weight functions based on the
physics associated with the problem:

\emph{One-temperature theory:} the obvious choice for the temperature of the Maxwellian
weighting function is the neutral gas temperature \cite{Kihara1953, 
  Kumar1973}. This choice includes, as a special case,
the Chapman-Enskog theory.

\emph{Two-temperature theory:} by allowing the temperature of the weighting
function to differ from the netural gas temperature, a more efficient basis set
can be constructed. Kelly applied this to an ``effective''
electron temperature, Viehland and Mason \cite{Viehland1975} used the actual
ion temperature whereas Lin~\etal\ \cite{Lin1979} used this
temperature as a variable parameter to optimize the convergence of the basis.

\emph{Three-temperature theory:} to provide even more flexibility in the basis,
Lin~\etal\ 
\cite{Lin1979} used a drifted Maxwellian weighting function that included a
different temperature parallel and perpendicular to the field.

\emph{Bi- and multi-modal theories:} in order to approach distributions which
strongly deviate from Maxwellian-like behavior, several groups
\cite{Ness1990, White1999} attempted to
split the distribution into overlapping Maxwellian distributions used to
characterize different regions of the velocity distribution. Although this
causes the basis to become non-orthogonal, with an orthogonalisation procedure
required, good fits were obtained to x distributions. However, these
methods unfortunately require a great deal of careful selection and optimization
of the basis.

\emph{Gram-Charlier theory:} to extend the three-temperature theory,
Viehland~\cite{Viehland1994} incorporated skew and kurtosis in the weighting
function.

% \begin{enumerate}
% \item One-temperature theory:  a Maxwellian weight function at 
% the neutral gas temperature (includes as a special case the 
% Chapman-Enskog theory) (Kihara (1953); Mason and Schamp (1958); Kumar and Robson 
% (1973); Robson and Kumar (1973); Robson (1973)).
% \item Two-temperature theory:  a Maxwellian weight function at 
% a basis temperature different from the neutral gas temperature. Kelly 
% (1960) employed an `effective' electron temperature; Viehland 
% and Mason (1975) employed the actual ion temperature; Lin {\em 
% et al.} (1979) left the basis temperature as an arbitrary parameter used 
% to optimize convergence.
% \item Three-temperature theory:  A drifted Maxwellian weight function 
% characterized by different temperatures parallel and 
% perpendicular to the field (Lin {\em et al.} (1979)).
% \item Bi and multi-modal theories: A weighted sum of two or more Maxwellians, each 
% characterizing different regions of the velocity distribution  (Knierim {\em et al.} (1981); 
% Ness and Viehland (1990), White et al. 1999)
% \item Gram-Charlier theory:  A three-temperature theory which 
% incorporates skew and kurtosis in the weight function
% Viehland (1994).
% \end{enumerate}
% The earlier literature is summarized in Mason and McDaniel 
% (1988).

The theory of Lin {\em et al.} \cite{Lin1979} is particularly important within
the context of electron swarms.  Their work unified electron and ion swarm
theories utilising the elegant and efficient irreducible tensor formalism of
Kumar \cite{Kumar1966a,Kumar1967} and provided a different
perspective of the two-temperature theory of Viehland and
Mason~\cite{Viehland1975}.  This work was extended by Robson and Ness
to produce a comprehensive multi-term treatment of
reactive steady-state d.c. electron (and ion, Ness and Robson (1989), White et
al 2001) swarms, a.c. electron and ion (White et al 1999, 2001), electric and
magnetic fields (Ness 1994, White et al 1994, Dujko et al. 2008).

The utility of these theories lies in the use of Talmi transformations to
facilitates a mass-ratio expansion of the collision operator in terms of
$m/(m+m_0)$. For light particles, the mass ratio is generally $m/m_0 < 10^{-4}$
which commonly allows an electron transport calculation to include only the
first order contribution. As the mass ratio increases, $m/(m+m_0) \rightarrow 1/2$ and 
more terms in the expansion must be retained to achieve convergence. For ions that
are heavier than the neutral case, a similar expansion in terms of $m_0/(m+m_0)$
can be made instead.

The various theories described above present a spectrum of approaches of
differing complexities. On the simple side, there may be one or even no free
parameters in the optimization of the basis set and the corresponding
calculations are straightforward to implement and can be run without much
supervision. However, the simple properties of the basis set are often unable to
accommodate non-Maxwellian behavior, which is especially important for ion
transport. On the complex side of the spectrum, the large number of parameters
available to optimize the basis set can allow the capture of pratically all
relevant velocity distributions. However, such basis sets can be very sensitive
to the choice of parameters, requiring careful selection and monitoring of the
basis set in order to achieve convergence.
Furthermore, unsafe choices of parmeters can lead to unstable numerical convergence.

In this article, our goal is to present a framework that allows any of the above
choices of basis set to be implemented. The basis set need not be orthogonal
and there is no need to evaluate matrix elements in the basis
analytically. This framework is also aimed at implementing more general basis
sets that allow for the flexibility of the theories mentioned
above with large numbers of optimizable parameters, while maintaining good
convergence properties.

\section{Theory}
\label{sec:theory}

\subsection{The Boltzmann equation}

Let a gas mixture contain particles of type $i$, $i\in \{1,2\}$, with
electric charge $q_i$, mass $m_i$, and number-density $n_i$, where
hereafter the subscript $i=1$ is reserved for the swarm particles and
$i=2$ for the background gas particles.  The macroscopic collection of
the particles is described statistically by the homogenous phase-space
distribution functions $\widetilde{F}_i(\vecv)$ and $F_i(\vecv)$
normalized by
\beq \widetilde{F}_i(\vecv)\equiv n_i F_i(\vecv), \ \ \ \int d
\vecv \ F_i(\vecv) = 1.
\label{f_norm}
\eeq
The background gas is assumed to be in equilibrium at a temperature $T_2$,
\beq
F_2(\vecv) = W_M(\alpha_2,\vecv), \ \ \ \alpha_2^2 = m_2/(k T_2),
\label{f_2_F}
\eeq
where $W_M(\alpha,\vecv)$ denotes the normalized Maxwell distribution
\[
W_M(\alpha, \vecv) =  \alpha^3 (2\pi)^{-3/2}  e^{-\alpha^2 v^2/2}.
%\ \ \int d\vecv  W_M(\alpha, v)  = 1.
\]
%% The {\em swarm} condition could be summarized by \cite{RobRobson2006,LRM78}
%% \[
%% n_1 \ll n_2.
%% \]
Then within the swarm limit ($n_1 \ll n_2$ \cite{Robson2006,Lin1979a})
and in the presence of spatially uniform/homogeneous electric field
$\vecE$, the relevant time-independent Boltzmann equation becomes
%\cite{ChapmanCowlingEd2,LP81v10,RobRobson2006}
\cite{Chapman1970,Pitchford1981,Robson2006}
\[
\frac{q_1}{m_1} \vecE \cdot \frac{\partial \widetilde{F}_1(\vecv_1)}{\partial \vecv_1}
= J(\widetilde{F}_1, \widetilde{F}_2, \vecv_1),
\]
%which is converted to the normalized distributions (divided by $n_1 n_2$) to become
or alternatively,
\beq
\frac{q_1 \vecE}{m_1 n_2} \cdot \frac{\partial F_1(\vecv_1)}{\partial \vecv_1} = J(F_1, F_2, \vecv_1),
\label{BE_2}
\eeq
where only the ground state of the background gas of neutral ($q_2 = 0$) particles is considered, and
$J(F_1,F_2, \vecv_1)$ is known as the pair-wise Boltzmann collision term.

\subsection{Elastic scattering}

All pair-wise collisions are assumed to be localized (position
$\vecr$) and instant (time $t$).  Before an individual collision, an
electron and gas particle are described by their velocities $\vecv_1$
and $\vecv_2$ in the laboratory coordinate frame of reference (the $L$
system).  After the collision the corresponding velocities become
$\vecv_1'$ and $\vecv_2'$ within the $L$ system.  The pair-wise
interaction conserves the total momentum of the colliding system
$\vecP$, and hence the velocity of the center-of-mass $\vecG=\vecP/M$,
where the full set of relevant velocities and momenta are
parameterized as follows \cite{Chapman1970,Robson2006}
\[
M= m_1 + m_2, \ \ \ \mu_{12}=m_1 m_2/ M,
\]
\[
\mu_1 = m_1/M, \ \ \ \mu_2 = m_2/M,
\]
\[
\vecP = \vecP', \  \vecP=m_1 \vecv_1 + m_2 \vecv_2, \  \vecP'=m_1\vecv_1' + m_2\vecv_2',
\]
\[
\vecG = \vecG', \  \ \vecG=\mu_1 \vecv_1 + \mu_2 \vecv_2, \  \ \vecG'=\mu_1\vecv_1' + \mu_2\vecv_2'.
\]
The required transition $(\vecv_1, \vecv_2)\rightarrow (\vecv_1', \vecv_2')$ in the $\vecv$-variables can be parameterized
in terms of the $(\vecG, \vecg)\rightarrow (\vecG', \vecg')$ variables,
\[
\vecv_1 = \vecG + \mu_2 \vecg,\ \  \vecv_2 = \vecG - \mu_1 \vecg,
\ \  \vecg = \vecv_1 - \vecv_2,
\]
\[
\vecv_1' = \vecG + \mu_2 \vecg',\ \  \vecv_2' = \vecG - \mu_1 \vecg',
\ \  \vecg' = \vecv_1' - \vecv_2',
\]
\beq
\vecv_1' - \vecv_1 =  \mu_2 (\vecg' - \vecg),
\ \ \ \vecv_2' - \vecv_2 =  \mu_1 (\vecg' - \vecg). \label{vp1_v1}
\eeq

The elastic collision rotates the relative collision velocity $\vecg$ into $\vecg'$,
\[
g=|\vecg| = |\vecg'|, \ \ \vecg = g \hat{\vecg}, \ \ \ \vecg' = g \hat{\vecg}'
\]
where $\vechg$ and $\vechg'$ are the unit vectors along $\vecg$ and $\vecg^\prime$.
The $\vecg \rightarrow \vecg'$ transition is
described by the  differential cross section
$\sigma(\vecg', \vecg)=\sigma(g, \chi)$, $\cos\chi = \vechg' \cdot \vechg$, such that
the number of electrons
scattered into the solid angle element $ d\vechg'$
per unit time (and per single collision) is given by
\beq
g \sigma(g, \chi)\ d\vechg'.
\eeq
% where the prefactor of $g$ side 
% the flux of swarm particles (or the current density) near a single gas particle.
The collision term $J(F_1,F_2,\vecv_1)$ is a function of $\vecv_1$,
\beq
J(F_1, F_2, \vecv_1) = G(F_1, F_2, \vecv_1) - L(F_1, F_2, \vecv_1),
\label{J_v2_def}
\eeq
\[
G(F_1, F_2, \vecv_1) = \int  d\vecv_2 d\vechg'  \ g \sigma(g\chi) \ F'_1 F'_2,
\]
\[
L(F_1, F_2, \vecv_1) = \int d\vecv_2 d\vechg'  \ g \sigma(g\chi) \ F_1 F_2,
\]
\[
F_i \equiv F_i(\vecv_i), \ \ \ \ F'_i \equiv F_i(\vecv'_i),
\]
where the gain $(G)$ and loss $(L)$ contributions are shown separately.

Other processes, such as inelastic scattering, can be included into
$J(F_1,F_2,\vecv_1)$ either exactly or in a simplified form. However, in this
paper, we focus only on elastic scattering, although the framework is unchanged
by the addition of other processes.

\subsection{Dimensional scaling}

Let $\vecu$-vectors denote dimensionless equivalent of the $\vecv$-vectors via
\beq
\vecv_i = \vecu_i / \alpha_i,\ \ \
\vecv'_i  = \vecu'_i / \alpha_i,\ \ \
\alpha_i^2 = m_i/(k T_i),
\label{u_alpha_v}
\eeq
where $i\in \{1,2\}$.
The $F_i(\vecv_i)$ distributions are then converted to $f_i(\vecu_i)$ distributions via
\beq
F_1(\vecv_1) = F_1(\vecu_1/\alpha_1) \equiv \alpha_1^3 f_1(\vecu_1),
\eeq
\[
%F_2(\vecv_2) =  W_\mathrm{M}(\alpha_2, v_2) = \alpha_2^3 w_\mathrm{M}(u_2), \ \ \  f_2(\vecu_2) \equiv w_\mathrm{M}(u_2),
F_2(\vecv_2) = \alpha_2^3 w_\mathrm{M}(u_2), \ \ \  f_2(\vecu_2) \equiv w_\mathrm{M}(u_2),
\]
\begin{equation}
  \label{eq:scaled_weighting}
w_\mathrm{M}(u) =  (2\pi)^{-3/2}  \exp(-u^2/2),
\end{equation}
\beq
\int d\vecu \ f_i(\vecu)  = 1, \ \ \ \int d\vecu \ w_\mathrm{M}(u)  = 1.
\label{f1_norm}
\eeq
To clarify the notation, once $f_1(\vecu_1)$ is found, $F_1(\vecv_1)$ can be reconstructed via
\beq
F_1(\vecv_1)  = \alpha_1^3 f_1\left( \alpha_1 \vecv_1 \right).
\eeq

The collision term $J(F,F_2,\vecv)$ (\ref{J_v2_def}) is a function of
$\vecv \equiv \vecv_1$ and hence it could also be converted to 
$\vecu$-space via Eqs.~(\ref{u_alpha_v}) and (\ref{f1_norm}),
\[
J(F, F_2, \vecv) \equiv \alpha_1^3 J(f, f_2,\vecu),
\]
\begin{equation}
  \label{eq:COLLINT}
J(f, f_2, \vecu) = G(f_1, f_2, \vecu) - L(f_1, f_2, \vecu),
\end{equation}
\[
G(f, f_2, \vecu) = \int  d\vecu_2 d\vechg'  \ g \sigma(g\chi) \ f'_1 f'_2,
 \]
\[
L(f, f_2, \vecu) = \int d\vecu_2 d\vechg'  \ g \sigma(g\chi) \ f_1 f_2,
 \]
\[
f_i \equiv f_i(\vecu_i), \ \ \ f'_i \equiv f_i(\vecu'_i),
\ \ \ \vecu \equiv \vecu_1, \ \ \  \vecu' \equiv \vecu'_1,
\]
arriving at the dimensionless version of the Boltzmann Eq.~(\ref{BE_2}),
\beq
\left(\frac{q_1}{k T_1 \tilde{\sigma} }\right) \frac{\vecE}{n_2}
\cdot \frac{\partial f(\vecu)}{\partial \vecu} =
\frac{\alpha_1}{\tilde{\sigma}} J(f, f_2, \vecu),
\label{BE_3}
\eeq
%% where both sides are multiplied by $\alpha_b / a_0^2$ to remove the units of $[length]^3[time]^{-1}$,
%% and where $\vecE/n_2$ is traditionally specified in the units of Townsend (Td),
%% and $a_0$ is Bohr radius (see Eq.~\ref{AU}).
where both sides have been multiplied by $\alpha_1 / \tilde{\sigma}$ and
$\tilde{\sigma}$ is the unit for cross section. Note that
$\vecE/n_2$ is traditionally specified in the units of Townsend (Td).
%% , and $a_0$ is Bohr radius (see Eq.~\ref{AU}).

\subsection{Unified expansion}

While $f_2(\vecu)$ is fixed, $f_1(\vecu)$ is expanded via
% the spherical
% harmonics $Y_{lm}(\vechu)$ and the arbitrary basis $R_{nl}(u)$ which is
% orthonormal with respect to the weighting function $w(u)$,
an orthonormal basis $\psi_\gamma(\vecu)$ with respect to the weighting function $w(\vecu)$,
\beq f(\vecu) = \sum_{\gamma}
f_{\gamma} w(u) \psi^*_{\gamma}(\vecu), \ \ f_{\gamma} = \int d\vecu \ \psi_\gamma(\vecu) f(\vecu),
\label{f1_sum}
\eeq
% \beq
% \psi_\gamma(\vecu) \equiv R_{nl}(u) Y_{lm}(\vechu),
% \ \ \ \gamma \equiv (n,l,m),
% \label{gamma}
% \eeq
%% \beq
%% \int d\vecu \ w(u) \ \psi_\gamma(\vecu) \psi_{\gamma'}^*(\vecu) = \delta_{\gamma \gamma'}.
%% \label{Burnett_orth}
%% \eeq
Hereafter the charged swarm particle subscript $i=1$ will be suppressed where possible,
\[F(\vecv)\equiv F_1(\vecv_1), \ \ \ \ f(\vecu)\equiv f_1(\vecu_1).
\]
As the basis is orthogonal, we require:
% \[
% \int u^2 du \ w(u) \ R_{nl}(u) R_{n'l}(u) = \delta_{nn'},
% \]
% \[
% \int d\vechu \ Y_{lm}(\vechu) Y_{l'm'}^*(\vechu) = \delta_{ll'} \delta_{mm'}.
% \]
\beq
\int d\vecu \, w(\vecu) \psi_\gamma \psi^*_\gamma = \delta_{\gamma\gamma^\prime}
\eeq

%% Only the $m=0$ subset of $\{Y_{lm}\}$ is actually required
%% for the considered case of spatially-homogeneous constant electric field.

Since $J(f, f_2, \vecu)$ is also a function of $\vecu$, it is expanded via $\{\psi_\gamma\}$
as per Eq.~(\ref{f1_sum}),
\[
J(f, f_2, \vecu) =  \sum_{\gamma} J_{\gamma}(f,f_2) w(u) \psi^*_{\gamma}(\vecu),\\
\]
\[
J_{\gamma}(f, f_2) \equiv \int d\vecu \ J(f, f_2, \vecu)  \psi_\gamma(\vecu).
\]
The final expressions for the collision matrix
$J_{\gamma \gamma'}$ are obtained by substituting expansion (\ref{f1_sum}) into the collision integral (\ref{eq:COLLINT})
arriving at
\beq
J_{\gamma}(f, f_2) = \sum_{\gamma'} J_{\gamma \gamma'} f_{\gamma'}, \ \ \ 
J_{\gamma\gamma'} = G_{\gamma\gamma'} - L_{\gamma\gamma'},
\label{J_gg}
\eeq
\beq
G_{\gamma\gamma'} = \int d\vecu d\vecu_2 d\vechg' \ w(\vecu)  \ g \sigma(g,\chi) \
\psi^*_{\gamma'}(\vecu') f'_2 \psi_{\gamma}(\vecu)  ,
\label{G_gg}
\eeq
\beq
L_{\gamma\gamma'} = \int d\vecu d\vecu_2 d\vechg' \ w(\vecu) \ g \sigma(g,\chi) \
\psi^*_{\gamma'}(\vecu) f_2 \psi_{\gamma}(\vecu).
\label{L_gg}
\eeq

The left hand side of Eq.~(\ref{BE_3}) is dealt with in a similar fashion arriving
at the electric field interaction matrix
\beq
E_{\gamma \gamma'} = \int d\vecu \ w(\vecu) \ \psi_\gamma(\vecu) \frac{\partial \psi^*_{\gamma'}(\vecu) }{\partial u_z},
\label{E_gg}
\eeq
where the $z$-axis is chosen to aline with the external electric field $\vecE$
in the same or opposite direction for positive or negative ions, respectively.
For particular cases, (including the Burnett functions considered later), the
preceding matrix element of the gradient operator has a known standard solution,
see \S 5.7 of \citet{Edmonds1974}.

In general, we calculate the multidimensional integrals $G_{\gamma \gamma'}$,
$L_{\gamma \gamma'}$ and $E_{\gamma\gamma^\prime }$ numerically. Gauss-Hermite quadratures
\cite{NIST:DLMF} are used for the integrals over the magnitudes $u_i$,
$i\in\{1,2\}$, Lebedev quadratures are used for the angular integrations
$\vechu_1$ and $\vechg'$ and a Gauss-Legendre quadrature is used for
$\cos\theta_2=\vechu_2\cdot\vechz$. This allows for an arbitrary choice of basis function
for $\psi_\gamma(\vecu)$ to be used.

Combining the $E_{\gamma\gamma'}$ and $J_{\gamma\gamma'}$ matrices, the final
equation for the $\{f_{\gamma}\}$-coefficients (\ref{f1_sum}) becomes
\beq
\sum_{\gamma'=0}^{N_B} A_{\gamma\gamma'} f_{\gamma'}=0, \ \ \ \
A_{\gamma\gamma'} = \zeta_{12} E_{\gamma\gamma'} - J_{\gamma\gamma'},
\label{sum_A_f}\eeq
\[
\zeta_{12}= q_1 E/(kT_1 n_2),\ \ \ 0 \le \gamma \le N_B.
\]
where $N_B$ is the number of basis functions.

The normalization of $f(\vecu)$ in \eqref{f1_norm} can be used to replace
one of the equations represented in \eqref{sum_A_f} with the
normalization condition:
\beq
\sum_{\gamma'=0}^{N_B} \xi_{\gamma'} f_{\gamma'}= 1, \ \ \ \
\eeq
where $\xi_\gamma = \int d\vecu\ \psi_\gamma(\vecu)$ are the weights of the numerical
integration over the distribution.

\subsection{Choice of basis}
Many choices for $\psi_\gamma(\vecu)$ are available to provide an implementation for the
framework and allow for the numerical solution to~\eqref{f1_sum}. These basis
functions must be well-suited to the physical system in order to obtain a
reasonably accurate solution after a truncation to a computationally feasible
size of basis. Careful consideration should therefore be given to the asymptotic
behaviour for $|u|\rightarrow \infty$ and $|u|\rightarrow 0$, as well as to any sharp profiles that may
be present in the collision operator.

It is common to use a spherical harmonic expansion to represent the angular part
of the basis:
\beq
\psi_\gamma(\vecu) \equiv R_{nl}(u) Y_{lm}(\vechu), \quad \gamma \equiv (n,l,m).
\eeq
where we may restrict the set to $m=0$ for the case of a constant electric field
aligned along the $z$-axis. We point out that this is simply a common choice for
electrons that is well-suited to near-thermal distributions but it is not
necessary in the framework presented here. In the spherical harmonic expansion,
the matrix elements $E_{\gamma\gamma^\prime}$ of the field term can be obtained easily by
converting the partial $u_z$-derivative:
\[
 \frac{\partial }{\partial u_z} = \cos\theta \frac{\partial }{\partial u} - \frac{\sin \theta}{u} \frac{\partial }{\partial \theta}.
\]
Consequently, for $m=m'=0$
\[
E_{\gamma \gamma'} = a_{ll'} R^{nn'}_{ll'},
\]
where the only nonzero angular matrix elements $a_{ll'}$ are
\[
a_{l+1,l}=(l+1) [(2l+1)(2l+3)]^{-1/2},
\]
\[
a_{l-1,l}=l  [(2l-1)(2l+1)]^{-1/2}.
\]
The corresponding radial contributions are given by
\[
R^{nn'}_{l+1,l}= \int u^2 du\  w(u) R_{n,l+1}
\left( \frac{\partial }{\partial u} - \frac{l}{u} \right) R_{n'l},
\]
\[
R^{nn'}_{l-1,l} =  \int u^2 du\  w(u) R_{n,l-1}
\left( \frac{\partial }{\partial u} + \frac{l+1}{u} \right) R_{n'l},
\]

We must now specify $\{R_{nl}\}$ in order to obtain expressions for
$R^{nn^\prime}_{ll^\prime }$ and $J_{\gamma\gamma^\prime}$. A reasonable choice is for a basis that is
well-suited to thermal distributions. By choosing the weighting function,
\begin{equation}
w(u) = w_M(u)
\end{equation}
from eqn~\eqref{eq:scaled_weighting}, we are able to select a basis
temperature $T_1$,
% which is present in $\alpha_1$ of eqn~\eqref{u_alpha_v},
to optimise the choice of basis function. We then take the speed
functions to be
\[
R_{nl}(u) = C_{nl}  r^{l}L_{n}^{(l+1/2)}(r^2),
\]
where $L_n^{(a)}$ are the Laguerre polynomials \cite{NIST:DLMF}, $r=
u/\sqrt{2}$ and 
\[
C_{nl}^2 =
2 \pi^{3/2} \left( n!\right) / \Gamma \left( n+l + 3/2\right).
\]
This set of $\{\psi_\gamma \}$-functions is known as the Burnett functions \cite{Kumar1966,White2009a}.

For the case that we are considering, only the $m=0$ subset of $\{Y_{lm}\}$ is
actually required due to the spatially-homogeneous constant electric field.  This
subset is easily defined using purely-real functions \cite{Edmonds1974},
\[
Y_{l,m=0}(\theta, \varphi) = \sqrt{(2l+1)/(4\pi)} \ P_l(\cos \theta),
\]
where $P_l$ are the Legendre polynomials \cite{NIST:DLMF}.

The following are the first few basis functions,
which are directly linked to the transport properties of the swarm particles,
\[
\psi_{000}(\vecu) =1,\ \ \ \psi_{100}(\vecu) =(3-u^2)/\sqrt{6},
\]
\[
\psi_{010}(\vecu) = u_z,\ \ \ \psi_{020}(\vecu) =(3u^2_z-u^2)/\sqrt{12},
\]
\beq
u^2 =3  - \sqrt{6} \ \psi_{100}(\vecu),  \ \ \ u_z = \psi_{010}(\vecu),
 \label{u2_2}
\eeq
\beq
u_z^2 =1  + \left[2 \psi_{020}(\vecu) - \sqrt{2} \ \psi_{100}(\vecu)\right] / \sqrt{3}.
 \label{uz_2}
\eeq
Note that the non-normalized Burnett functions are also used in the literature,
for example \cite{Lin1979a,Mason1988} utilized
\[
\psi^{\mbox{\tiny MM}}_{nl0}(\vecu) = r^{l}L_{n}^{(l+1/2)}(r^2) w(u) P_l(\cos \theta),
\]
where $r= u/\sqrt{2}$, and only $m=0$ is shown.

The only constraint on the size of the basis is the sensitivity of the
integrations over the basis functions. In the case of the Burnett functions used
here, very large values of N give rise to a loss of orthogonality when
performing the numerical integrations for $J_{\gamma\gamma^\prime}$ and
$E_{\gamma\gamma^\prime}$. In this case the expansion of the distribution function begins to
break down. In this article, we take enough quadrature points to ensure that the
orthonormality of the basis functions is to within $10^{-11}$, which allows for
up to $N=40$ Laugerre polynomials to be used.

\subsection{Transport properties}
\label{sec:transport-properties}

When considering ions as the swarm particles, the following transport properties
are of interest \cite{Mason1988}:
% mean ion temperature $(T_\mathrm{ion})$;
mean ion energy $(\varepsilon)$, drift velocity $(W)$
and transverse and longitudinal temperature tensor elements ($T_T$ and $T_L$)
% and diffusion coefficients ($nD_T$ and $nD_L$)
defined as
\beq
\varepsilon = \frac{1}{2} m \langle v^2 \rangle,
%\ \ \ \frac{3}{2} k T_\mathrm{ion} = \frac{1}{2} m \langle v^2 \rangle,
\label{def_eps}
\eeq
\beq
W = \langle v_z \rangle,
\ \ \ \varepsilon_z = m \langle  v_z^2  \rangle / 2,
\label{def_w} \eeq
\beq
k T_T = m \langle v^2_x \rangle  = m \langle  v_y^2  \rangle=\varepsilon - \varepsilon_z,
\label{T_T} \eeq
\beq
k T_L = m \left[ \langle v^2_z \rangle -  \langle v_z \rangle^2 \right]= 2\varepsilon_z - mW^2,
\label{T_L} \eeq
\beq
\langle P \rangle \equiv \int d\vecv P(\vecv) F(\vecv),
\eeq
where $m \equiv M_1$.
%and $n \equiv n_2$.

For the Burnett functions, the transport properties can be easily expressed in
terms of the first few expansion coefficients $f_{\gamma}$, see Eqs.~(\ref{f1_sum}),
(\ref{u2_2}), and (\ref{uz_2}),
\begin{equation}
  \label{eq:Burnett-W}
W = f_{010} / \alpha_1.
\end{equation}
\begin{equation}
  \label{eq:Burnett-eps}
\varepsilon = \frac{3 kT_b}{2} \left[ 1-\sqrt{\frac{2}{3}} f_{100} \right],
\end{equation}
\begin{equation}
  \label{eq:Burnett-epsz}
\varepsilon_z = \frac{kT_b}{2} \left[ 1 + \frac{ 2 f_{020}-\sqrt{2} f_{100}}{\sqrt{3}}\right].
\end{equation}

\section{Results and discussion}
\label{sec:results}

\begin{table*}[tb]
%\begin{tabularx}{\textwidth}{l l l l l l l l l l l l}
%\begin{tabularx}{\textwidth}{. . . . . . . . . . . .}
%\begin{tabularx}{\textwidth}{c c c c . . . . . . . .}
%\begin{tabularx}{\textwidth}{c c c c . . . . .}
\begin{tabular}{*{4}{c} *{8}{.}}
\hline
%\mc{$m/m_0$} & \mc{$E/n_2$} & \mc{$T$} & \mc{L} & \mc{N} & \mc{$\varepsilon$} & \mc{$W$} &  \mc{$nD_L$}  & \mc{$nD_T$}  & \mc{$T_T$} & \mc{$T_L$} & \mc{-$n\gamma$}  \\
% & \mc{(Td)} & \mc{(K)} & & & \mc{($10^{-2}$ eV)} & \mc{(10$^3$ ms$^{-1})$} & \mc{($\mathrm{10^{22}\,m^{-1}s^{-1}}$)} & \mc{($\mathrm{10^{22}\,m^{-1}s^{-1}}$)} & \mc{(10$^2$ K)} & \mc{(10$^2$ K)} & \mc{(10$^{18}$ kg s$^{-2}$)}
\mc{$m/m_0$} & \mc{$E/n_2$} & \mc{$T$} & \mc{L} & \mc{N} & \mc{$\varepsilon$} & \mc{$W$} &
\mc{$T_T$} & \mc{$T_L$} & \mc{$\Delta f_0$} & \mc{$ \Delta f_1 $} & \mc{$\Delta f_2$} \\
 & \mc{(Td)} & \mc{(K)} & & & \mc{($10^{-2}$ eV)} & \mc{(10$^3$ ms$^{-1})$} &
\mc{(10$^2$ K)} & \mc{(10$^2$ K)} & & & 
\input{table_data.tex}
                                        \hline
\end{tabular}
\caption{\label{tab:transport-quantities}Converged values for the transport
  quantities, as defined in section~\ref{sec:transport-properties}. Also given
  are the number of Legendre polynomials (L) and number of Sonine polynomials
  (N) used. For comparison, the differences to the results from MC simulations
  are included as bracketed values and the integrated absolute deviations are
  shown as $\Delta f_l$ see~\eqref{eq:distdiff}.
}
\end{table*}

\subsection{The model}

We apply the framework of section \ref{sec:theory} to find solutions of the
standard hard-sphere benchmark model. The cross-section of the hard-sphere model
is isotropic and given by $\sigma(\epsilon)=6\,\mathrm{\AA}$.  All numerical computations
are performed using atomic units (a.u.), which are defined by setting electron
mass $(m_e)$, absolute value of electron charge $(e)$, reduced Planck' constant
$(\hbar)$, and Coulomb's constant $1/(4\pi \epsilon_0)$ to unity.
% The following are some relevant conversions to and from a.u. as per \cite{CODATA2010}:
% \beq \barr{l}
% E_h$(a.u. Hartree)$=27.211385$~eV$,\\
% $(1~K)$k=8.6173324 \times 10^{-5}$~eV$,\\
% $(1~K)$k=3.1668114 \times 10^{-6}$~$E_h,\\
%   a_0=0.52917721$~\AA$, \ \ \
% a^2_0=0.28002852$~\AA$^2,\\
% a_0^2 E_h=7.6199638$~eV$\cdot$\AA$^2,\\
% $(1~Tb)$e=10^{-21}/(a_0^2 E_h) =0.013123422,
% \earr \label{AU} \eeq
% where $E_h$ is the unit of energy (Hartree), $a_0$ is the unit of length (Bohr radius),
% $k$ is the Boltzmann constant, \AA$^2=10^{-10}$m, and where
% the Townsend (Td) unit 1Td=$10^{-21}$~V$\cdot$m$^2$.
% For example, in Eq.~(\ref{BE_3}) the left-hand side gradient multiplier $eE/(kT_1a_0^2 n_2)=$ 2.0720245
% for $E/n_2=1$~Tb and $T_1=2000$~K.
We specify all electric fields in the ratio $E/n_2$ with units of Townsend
($\mathrm{1\,Td = 10^{-21}\,V\cdot m^2}$). On publication, the complete java source code will be released as an open-source via github.

We consider several different regimes of mass ratio,
$m/m_0 = \{10^{-4}, 0.1, 1, 2, 10\}$. The lowest mass ratio $m/m_0 = 10^{-4}$
corresponds to a light charged particle (e.g. electron) whereas the other ratios
correspond to ions.  For all regimes, we focus on convergence of the solution,
quantified by the transport quantities and the distribution function
$F_1(\vecv)$. These are compared with identical quantities sampled from
Monte-Carlo (MC) simulations, which have been extensively benchmarked
\cite{Tattersall2015} and include temperature effects of the neutral background
\cite{Ristivojevic2012}.

An appropriate choice of temperature for the basis functions is important to
achieve stable solutions. In particular, small variations of the basis
temperature should not break the convergence of the solution. After many
trials, we found that the following empirical choice for basis temperature worked well:
$T_b = x E/n_2 + T_2$, where $x=7000, 200, 100, 80, 100 \, \mathrm{K/Td}$ for
$m/m_0 = 10^{-4}, 0.1, 1, 2, 10$ respectively.

\subsection{Transport quantities}

For all mass ratios, we find that the transport quantities converge quickly to within
$0.1\%$ of the MC results, requiring at most only $L=5$ Legendre components and
$N=10$ Laguerre polynomials in the Burnett function expansion.
%In figure~\ref{fig:transquant-convergence} we show the convergence of mean energy and drift
%In figure~\ref{fig:combconv-T0} we show the convergence of the drift velocities for the different mass ratios at $T=0$.
Although the higher mass ratios are the slowest to converge,
% (addressed in more detail in section \ref{sec:results-distfuncs}),
their computational time is negligible.
%The case of $T=293$~K fares even better than $T=0$ are can be seen in
%figure~\ref{fig:combconv-T293}.
In table~\ref{tab:transport-quantities} we show the converged values for all
cases, compared with the same quantities from the MC simulations.

The rapid convergence of the transport quantities can easily be understood
through equations~\eqref{eq:Burnett-W}--\eqref{eq:Burnett-epsz}. Each of the transport
quantities depends on only one or two of the expansion
coefficients and the strength of coupling of these coefficients to higher basis
functions decreases. Hence, as the truncation in the Burnett expansion is extended,
the additional coefficients have a negligible effect on the transport quantities.

We note that the convergence of the transport quantities can occur even when the
reconstructed distribution function is significantly different from the MC
result, which we discuss below in further detail.

\subsection{Distribution function}
\label{sec:results-distfuncs}

The importance of obtaining correct distribution functions, especially in the
low-energy and tail regions of the distribution, is likely to be missed if one
only considers transport quantities. There are many scenarios which require
accurate distribution functions. For example, some loss processes
(e.g. recombination and attachment for electrons, or direct annihilation for
positrons) are very strong and sharply peaked for low energies and so it is
essential to obtain correct low energy behaviour of the distribution
function. Alternatively, ``soft'' potentials can cause ``run-away'' effects in
the tail of the distribution, and these high-energy particles then strongly skew
the distribution averages.

We demonstrate the convergence and agreement between the MC results and the
Burnett expansion by plotting the expansion of the distribution in Legendre
components, i.e.
\begin{equation}
  f(\vecv) = \sum_l f_l(v) P_l(\cos \theta).
\end{equation}
To obtain a quantitative value to represent the convergence of the distribution
functions, we also calculate
\begin{equation}
  \Delta f_l = \int v^2 dv \, |f_l^\mathrm{Boltz}(v) - f_l^\mathrm{MC}(v)|,
  \label{eq:distdiff}
\end{equation}
which is an error that is relative to the normalisation of $f(\vecv)$, i.e. $\int
v^2 dv\, |f_0(v)| \equiv 1$.

\begin{figure}[tb]
  \begin{center}
  \includegraphics[width=0.8\linewidth]{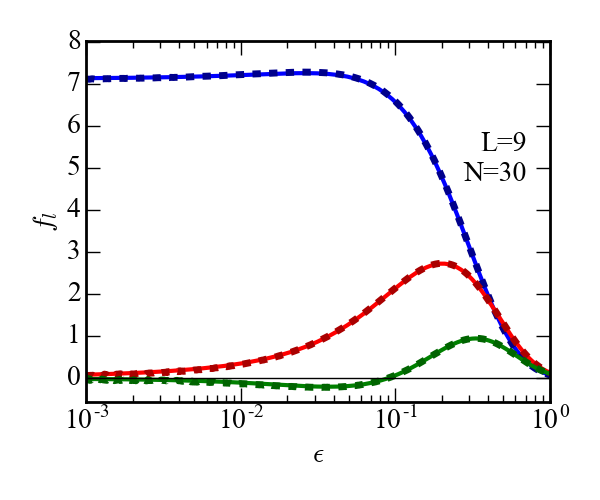}
  \end{center}
  \caption{\label{fig:mratio-0o1-dists} The best estimate of the distribution
    function, plotted in Legendre components $f_l$ for $l=0$ (blue), $l=1$ (red)
    and $l=2$ (green) for $m/m_0=0.1$, $E/n_2=12$~Td and $T=293$~K. The result from a Monte-Carlo simulation
    (thick dashed lines) is shown for comparison.
    }
\end{figure}

\begin{figure}[tb]
  \begin{center}
  \includegraphics[width=0.8\linewidth]{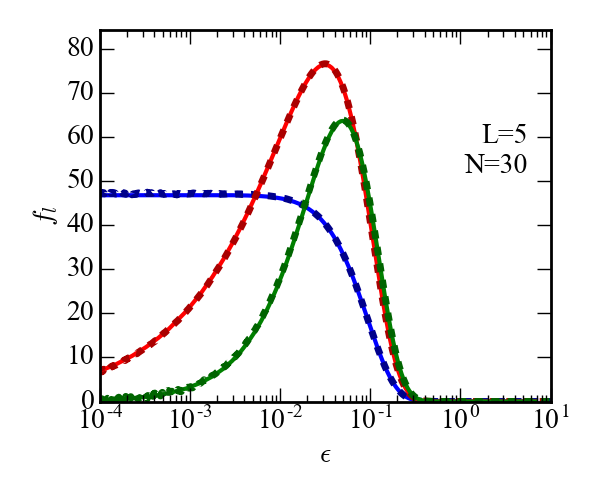}
  \includegraphics[width=0.8\linewidth]{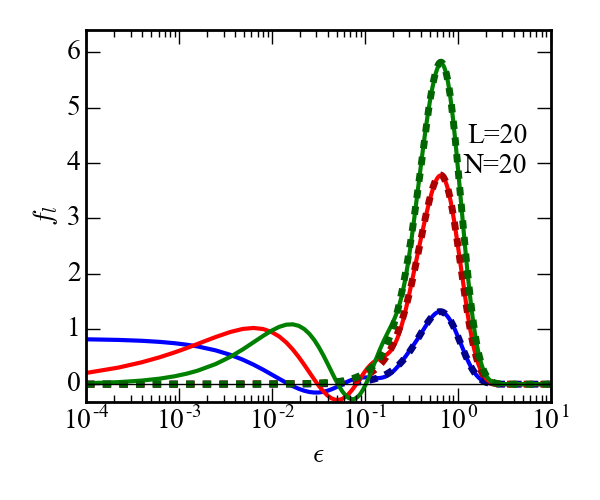}
  \end{center}
  \caption{\label{fig:mratio-10-dists} As in figure~\ref{fig:mratio-0o1-dists}
    but for $m/m_0=10$ and as a comparison between a) $E/n_2 = 2$~Td and b)
    $E/n_2 = 12$~Td. When the field is large, the distribution becomes highly
    non-thermal and the Burnett function expansion cannot reproduce the flat
    behaviour of the distribution at low energies.
    }
\end{figure}

\begin{figure}[tb]
  \begin{center}
  \includegraphics[width=0.8\linewidth]{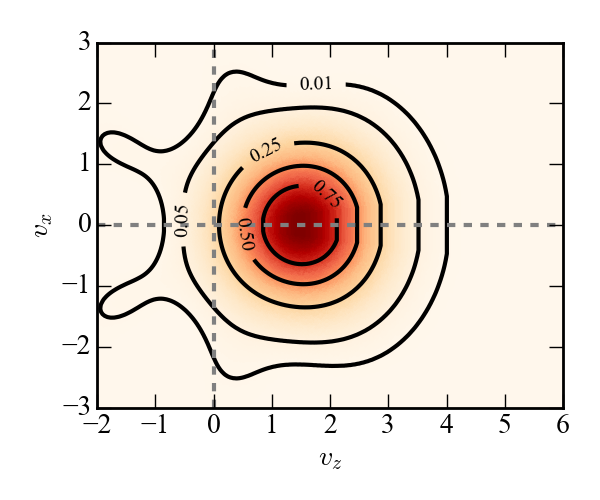}
  \includegraphics[width=0.8\linewidth]{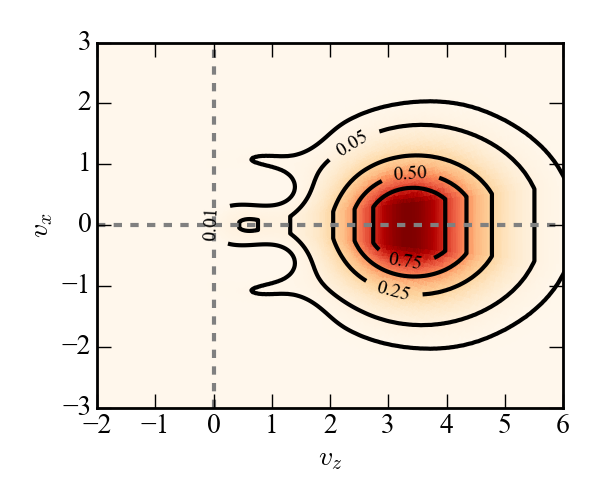}
  \end{center}
  \caption{\label{fig:3dplots} Surface plots of the distribution, $f(v_x,0,v_z)$ for
    $m/m_0=10$ and $T=293$~K. In a) $E/n_2 = 2$~Td and the distribution has a
    resonably strong presence around the origin. In b)
    $E/n_2 = 12$~Td and the distribution has been significantly shifted away
    from the origin.
  }
\end{figure}

For mass ratios $m/m_0 \leq 1$, we obtain good agreement between the Boltzmann
and MC distribution functions for all fields that we investigated, see
figure~\ref{fig:mratio-0o1-dists}. The agreement between the two calculations
continues as the mass ratio becomes larger ($m/m_0 \leq 2$), however the
number of basis functions required to achieve convergence quickly grows. For high
fields the convergence in the distribution functions requires many more basis
functions than for convergence in the transport quantities alone.

For much larger mass ratios ($m/m_0 = 10$), the Burnett function expansion
begins to fail for higher fields, which can be seen in
figure~\ref{fig:mratio-10-dists}. In the two different fields shown, $E/n_2 = 2$
and $10$, it can be seen that the main difference in the distribution functions
lies in the suppression at low energies. This flat part of the distribution
cannot easily be reproduced by the Burnett functions and so the truncation
causes the solution to oscillate around in a manner similar to Gibbs phenomena
of a Fourier series expansion.

The failure of the Burnett expansion could be expected as the strong fields have
significantly distorted the distribution away from a Maxwellian distribution,
which is the preferred regime of the Burnett basis. In addition, the peak of the
distribution is strongly displaced from the origin, see
figure~\ref{fig:3dplots}, and a Legendre decomposition is not ideal for such
distributions. Hence, we believe that a new basis $\psi_\gamma(\vecu)$ is required,
which does not rely on a spherical harmonic expansion. It would also be
desirable for a different basis to more easily handle sharp features in the
cross section.

We note that we have also performed $T=0$ calculations which emphasize this
conclusion. For $m/m_0 < 1$ and $T=0$, the distributions remain similar enough
to a thermal distribution that the Burnett function expansion works well for all
field strengths. However, for equal mass ratios and larger, the distribution at
$T=0$ becomes highly non-thermal and the Burnett expansion fails for all field
strengths. This can be seen in the $\Delta f_l$ values shown in table~\ref{tab:transport-quantities}.

\begin{figure}[tb]
  \begin{center}
  \includegraphics[width=0.8\linewidth]{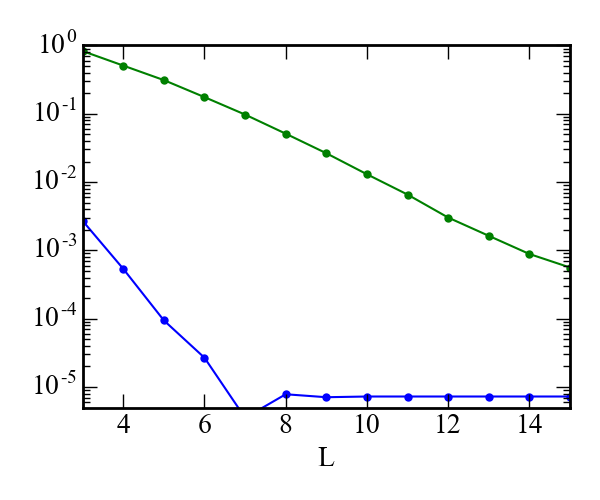}
  \end{center}

  \caption{\label{fig:convergence-trans-dist}Relative error convergence of drift velocity
    (blue line) and distribution function ($\Delta f_0$, green line), as the
    number of Legendre polynomials ($L$) is increased. The transport quantity
    converges much faster than the distribution.  Here, $T=293$~K, $m/m_0=10$ and
    $E/n_2=10$~Td. The relative error in the drift velocity is taken as the
    difference in values sampled from a MC simulation. A fixed value of
    $N=20$, the number of Laugerre polynomials, is used. Note that uncertainties in the MC
    results place a minimum on the calculate error in the drift velocity.
    }
\end{figure}

\subsection{Comments on benchmarking}
It is essential to test new code using benchmarks of simple models
before applying it to more complicated applications. However, the simple
comparisons in this paper highlight a subtlety that can easily give rise to
false confirmations for tests on code.

For example, even though the particular case of $m/m_0 = 10$ and $E/n_2 = 12$~Td
shown in this paper has clearly not reached convergence in the distribution
function, the corresponding transport quantities instead converge quickly. In
fact, the distribution has not only not converged, but is nonphysical, with
negative values appearing in the distribution. To highlight this, we show in
figure~\ref{fig:convergence-trans-dist}, the convergence for $E/n_2 = 10$~Td as
the truncation in number of Legendre polynomials in increased.

This indicates that benchmarking a transport code by obtaining agreement for
transport quantities only in simple models is not sufficient. Proper
consideration must be made for the distribution function, especially when an
application deviates significantly from the ideal and smooth case of common
benchmark models.

There are many examples of processes which may cause a code to break
down. Rather than list specific examples, it is instructive to see how an
incorrect distribution would affect the results of perturbation theory. Calculating the effect of a perturbation relies on accurate
distribution functions in the region where the new process is strongest, e.g. a
narrow window of energy, or for a particular pattern in angle. When the distribution function is unphysical, it could even be
possible to obtain an incorrect \emph{sign} of the perturbed quantity.

\section{Conclusion}

We have outlined a general framework to represent the
solution of Boltzmann's equation in the swarm limit, allowing arbitrary basis
functions and for arbitrary mass ratios. We have demonstrated how this framework
can be applied for the traditional basis set of Burnett functions. We obtained good
agreement between our Boltzmann solutions and an independent
Monte-Carlo simulation for a large range of parameters.

Although the Burnett function basis always converged quickly in the value of the
transport quantities, the distribution functions themselves were only accurately
represented for when they are sufficiently close to a thermal Maxwellian
distribution. This occurs for $m/m_0\geq1$ and large applied fields. The behaviour
of these functions can even be nonphysical, due to negative oscillations in
the solutions. This can cause some techniques to fail completely, such as in
perturbation theory which makes use of the distributions directly.

This suggests that accurate benchmarking requires models that qualitatively
capture the behaviour present in the desired application. Additionally,
benchmarks should include the comparison of the complete distribution function.

As we have demonstrated that the spherical harmonic expansion is likely to fail
when the peak of the distribution is moved away from the origin, we believe that
alternative basis sets should be used. In future work, we will apply and
contrast several different basis sets using the framework of
section~\ref{sec:theory}. In particular, we expect that a basis of B-splines
will provide a great deal of flexibility that allows for solutions in cases
which have strongly non-thermal distributions.

\bibliographystyle{apsrev}
%\bibliography{../qm_references}
\bibliography{ArbMass,library_RW_changed}

\end{document}

%% file: table_data.tex
\\ 
\hline
$10^{-4}$ & $0.1$& $0$ & $5$ & $30$ & 7.1182 & 1.7993 & 5.5060 & 5.5071 & 0.0006 & 0.0000 & 0.0001\\ 
 \emptycell & \emptycell & \emptycell & \emptycell & \emptycell & (-0.0014) & (-0.0005) & (-0.0012) & (-0.0012)\\ 
 \emptycell & \emptycell & $293$ & $5$ & $30$ & 9.4704 & 1.5942 & 7.3260 & 7.3268 & 0.0005 & 0.0001 & 0.0001\\ 
 \emptycell & \emptycell & \emptycell & \emptycell & \emptycell & (-0.0019) & (-0.0001) & (-0.0015) & (-0.0015)\\ 
\hline
$0.1$ & $2$& $0$ & $10$ & $30$ & 5.0023 & 1.4578 & 3.3407 & 3.9062 & 0.0005 & 0.0004 & 0.0005\\ 
 \emptycell & \emptycell & \emptycell & \emptycell & \emptycell & (0.0001) & (0.0000) & (0.0002) & (-0.0001)\\ 
 \emptycell & \emptycell & $293$ & $5$ & $30$ & 7.3312 & 1.2095 & 5.2959 & 5.7193 & 0.0006 & 0.0003 & 0.0006\\ 
 \emptycell & \emptycell & \emptycell & \emptycell & \emptycell & (0.0004) & (0.0001) & (0.0003) & (0.0003)\\ 
\hline
$1$ & $2$& $0$ & $20$ & $40$ & 3.9233 & 1.0283 & 1.1329 & 1.7529 & 0.0055 & 0.0225 & 0.0313\\ 
 \emptycell & \emptycell & \emptycell & \emptycell & \emptycell & (-0.0003) & (-0.0001) & (-0.0001) & (-0.0000)\\ 
 \emptycell & \emptycell & $293$ & $5$ & $30$ & 5.4262 & 0.6279 & 3.4156 & 3.8659 & 0.0006 & 0.0006 & 0.0007\\ 
 \emptycell & \emptycell & \emptycell & \emptycell & \emptycell & (0.0002) & (0.0000) & (0.0000) & (0.0003)\\ 
\hline
$2$ & $2$& $0$ & $15$ & $30$ & 5.4282 & 0.9813 & 0.9089 & 1.5151 & 0.0016 & 0.0150 & 0.0203\\ 
 \emptycell & \emptycell & \emptycell & \emptycell & \emptycell & (0.0002) & (0.0000) & (0.0001) & (0.0001)\\ 
 \emptycell & \emptycell & $293$ & $5$ & $30$ & 5.7019 & 0.5565 & 3.2762 & 3.7016 & 0.0005 & 0.0005 & 0.0008\\ 
 \emptycell & \emptycell & \emptycell & \emptycell & \emptycell & (-0.0003) & (-0.0000) & (-0.0003) & (-0.0001)\\ 
\hline
$10$ & $2$& $0$ & $15$ & $30$ & 18.6406 & 0.9178 & 0.7010 & 1.3335 & 0.0529 & 0.0129 & 0.1483\\ 
 \emptycell & \emptycell & \emptycell & \emptycell & \emptycell & (0.0004) & (-0.0000) & (0.0006) & (0.0001)\\ 
 \emptycell & \emptycell & $293$ & $5$ & $30$ & 9.1858 & 0.4874 & 3.1654 & 3.5615 & 0.0006 & 0.0016 & 0.0041\\ 
 \emptycell & \emptycell & \emptycell & \emptycell & \emptycell & (-0.0002) & (0.0000) & (-0.0007) & (-0.0001)\\ 